\def\dover#1#2{\hbox{${{\displaystyle#1 \vphantom{(} }\over{
   \displaystyle #2 \vphantom{(} }}$}}
\def\teq#1{$\, #1\,$}                         
\def\erg{\varepsilon}         
\def\pmb#1{\setbox0=\hbox{#1}%
  \kern-0.0125em\copy0\kern-\wd0
  \kern0.025em\copy0\kern-\wd0
  \kern-0.0125em\raise0.0433em\box0 }
\def\vol#1#2{$\;$ {\bf #1}, #2}                         
\def\reference{\noindent\hangindent=.5truecm\hangafter=1}
\def\apj{{\it Ap. J.}}

\def\aap{{\it Astron. Astr. }}   
\def\aaps{{\it Astron. Astr. Supp. }}

\documentstyle[twocolumn]{esa_sp_latex}

\begin{document}
%

\parindent 0pt
\parskip 10pt plus 1pt minus 1pt
\hoffset=-1.5truecm
\topmargin=-1.0cm
\textwidth 17.1truecm \columnsep 1truecm \columnseprule 0pt 

\input{psfig.tex}

\rm
\title{\bf GAMMA-RAYS FROM SUPERNOVA REMNANTS AND THE
          SIGNATURES OF DIFFUSIVE SHOCK ACCELERATION}

\author{{\bf Matthew G.~Baring,}$^{1\hbox{\dag}}$\ {\bf Donald C.~Ellison}$^2$ 
     {\bf and Isabelle Grenier}$^3$ \vspace{2mm} \\
$^1$Lab. for High Energy Astrophysics, NASA Goddard Space Flight Center, 
    Greenbelt, MD 20771, U.S.A. \\
$^2$Department of Physics, Box 8202, North Carolina State University, 
    Raleigh NC, 27695, U.S.A. \\
$^3$EUROPA-Universit\'e Paris, Paris VII and DAPNIA/Service d'Astrophysique,
    CE-Saclay, Gif/Yvette, France\vspace{2mm} \\
\dag Compton Fellow, Universities Space Research Association}

\maketitle

\begin{abstract}

While the definitive detection of gamma-rays from known supernova
remnants (SNRs) remains elusive, the collection of unidentified EGRET
sources that may be associated with SNRs has motivated recent modelling
of TeV emission from these sources.  Current theoretical models use
power-law shock-accelerated protons and electrons in their predictions
of expected gamma-ray TeV fluxes from those unidentified EGRET sources
with remnant associations.  In this paper, we explore a more detailed
non-linear shock acceleration model, which generates non-thermal proton
distributions and includes a self-consistent determination of shock
hydrodynamics.  We obtain gamma-ray spectra for SNRs allowing for the
cessation of acceleration to high energies that is due to the finite
ages and sizes of remnants.  Gamma-ray spectral cutoffs can be observed
in the TeV range for reasonable remnant parameters, and deviations
from power-law behaviour are found at all energies ranging from 1 MeV
up to the cutoff.  Correlated observations by INTEGRAL, Whipple and
other instruments may provide stringent constraints to our
understanding of supernova remnants.     \vspace {5pt} \\


\end{abstract}

\section{INTRODUCTION}
\label{sec:intro}

Supernova remnants have long been invoked as a principal source of
galactic cosmic rays, created via the process of diffusive Fermi
acceleration at their expanding shock fronts (e.g.  Drury 1983, Lagage
and Cesarsky 1983, Blandford and Eichler 1987).  Such systems can also
provide gamma-ray emission via the interaction of the cosmic ray
population with the ambient remnant environment; this concept was
explored recently by Drury, Aharonian and V\"olk (1994).  In their
model, the gamma-ray luminosity is spawned by collisions between the
cosmic rays and nuclei from the ambient SNR environment.  Inverse
Compton scattering involving shock-accelerated \teq{e^-} and the cosmic
microwave background and also IR/optical emission (from dust/starlight)
forms added components in the $\gamma$-ray SNR models of De Jager and
Mastichiadis (1996) and Gaisser, Protheroe and Stanev (1996).

Given the absence of definitive detections of gamma-rays from known
supernova remnants, the motivation for modelling these ``hypothetical''
sources hinges on over a dozen spatial associations of unidentified
EGRET sources at moderately low galactic latitudes with well-studied
radio SNRs.  These include IC 443, $\gamma$ Cygni and W44, and many
have neighbouring dense environments, which seem essential in order to
provide sufficient gamma-ray luminosity to exceed EGRET's sensitivity
threshold.  Confirmation of these associations will probably only come
with better gamma-ray angular resolution (e.g. at the soft gamma-ray
band addressed by the INTEGRAL mission) and/or positive TeV
detections.  The remnants associated with several of the EGRET
unidentified sources show an apparently low level of TeV emission, as
determined by Whipple (Lessard et al. 1995), which may be marginally
inconsistent with the spectra generated in the model of Drury,
Aharonian and V\"olk (1994); Gaisser, Protheroe and Stanev (1996) use a
slightly steeper accelerated proton distribution to provide greater
consistency between the EGRET and Whipple bands.  De Jager and
Mastichiadis (1996) suggest that there may be an intrinsic cutoff in
the SNR-generated cosmic ray population that would spawn a complete
absence of TeV emission, which is quite compatible with the data.

All of the above models invoke simple power-law accelerated particle
populations.  In this paper, we utilize the more sophisticated output
of shock acceleration simulations (e.g. Jones and Ellison 1991) to
address the issues of spectral curvature and the maximum energy of
acceleration in the context of SNR gamma-ray emission.  We use output
from the fully non-linear Monte Carlo simulations of Ellison and
Reynolds (1991) and Ellison, Baring and Jones (1996)
 to describe the accelerated population in environments where it
influences the dynamics of the SNR shell.  Such non-linear effects are
crucial to the modelling of gamma-ray emission in SNRs.  Our results
make clear predictions of what maximum energies of gamma-rays are
expected, and further make more accurate determinations of the
level of TeV emission relative to the MeV range in these sources.  We
find that the TeV/GeV flux ratio in our non-linear regime is reduced
by factors of a few below pure power-law proton scenarios, a
result that is very important for correlated observations by INTEGRAL,
Whipple and other gamma-ray instruments.

\section{FERMI ACCELERATION AT SHOCKS}
\label{sec:accel}

Diffusive shock acceleration is usually assumed to generate power-law
particle populations in astrophysical models.  This simple
approximation omits the effect the accelerated particles themselves
have on the hydrodynamics of their shocked environment.  Considerations
of such non-linear dynamics are essential to the gamma-ray SNR problem
since the peak luminosity arises at the onset of the Sedov phase
(Drury, Aharonian and V\"olk 1994) where the fraction of ram pressure
going into cosmic rays and nonlinear effects are maximized.  These
non-linear effects, well-documented in the reviews of Drury (1983) and
Jones and Ellison (1991), have a feedback on the acceleration mechanism
and its efficiency.  This becomes evident when it is observed that the
``test-particle'' power-law slope depends purely on the compression
ratio of flow speeds on either side of the shock, and that this ratio
is ultimately dependent on the hydrodynamics.

%
\vskip+0.5truecm
\centerline{\hskip -0.0truecm\psfig{figure=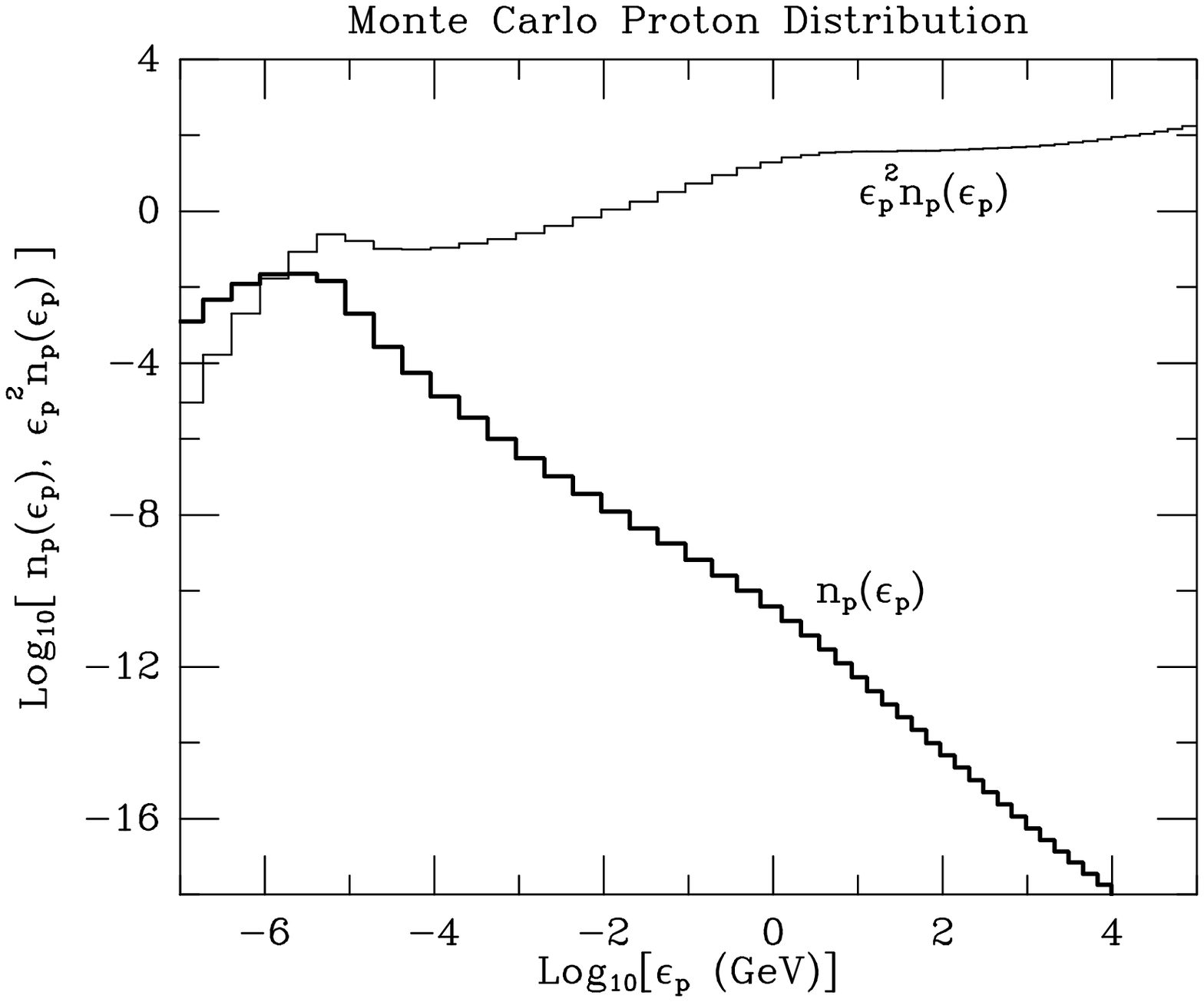,width=7.9truecm}}
\vskip-0.3truecm
{\smallskip \it Figure~1:  A typical proton distribution
\teq{n_p(\erg_p)}, of arbitrary normalization, resulting from Monte
Carlo simulations of particle acceleration at SNR shocks.  The
transition to relativistic energies produces a ``bump'' at around the
kinetic energy \teq{\erg_p\sim 1} GeV, a result of assuming a
momentum-dependent mean free path.  The \teq{\erg_p^2n_p(\erg_p)}
representation gives a good feel for the spectral curvature produced by
the non-linear hydrodynamics above 1 GeV.    \smallskip}

Our Monte Carlo simulation of Fermi acceleration determines the
acceleration efficiency and maximum energy of acceleration.  Although
steady-state, this method can roughly mimic the net features of the
time dependence of the SNR expansion and may describe the general
spectral properties quite well just prior to the start of the Sedov
phase.  A typical proton distribution is shown in Fig. 1.  Positive
curvature appears in the distribution above 1 GeV due to the non-linear
effects.  The difference in slopes about the proton rest mass energy at
\teq{\sim 1} GeV is just due to the varying dependence of momentum on
energy as prescribed by special relativity.  The pressure of the
accelerated population acts to slow down the fast-moving flow upstream
of the shock.

High-momentum protons have longer mean-free paths \teq{\lambda}, a
model assumption that is supported by observations made by plasma
experiments at the Earth's bow shock (e.g.  Ellison, M\"obius and
Paschmann 1990), and therefore typically influence the flow on larger
scalelengths.  The maximum compression ratio is achieved at the largest
scales, i.e. farthest from the downstream region (inside of the SNR
shell; see Ellison, Baring and Jones 1996 for typical velocity
profiles), implying an increased efficiency of accelerating
higher-energy particles relative to those at lower energies.  Upward
spectral curvature ensues, and this has a significant influence on
predictions of TeV fluxes in gamma-ray SNRs.  Non-linear predictions of
the spectral index differ significantly from the test-particle case
(Ellison, Baring and Jones 1996); in Fig.~1, a \teq{p^{-2}} proton
momentum distribution from a strong shock would be flat (i.e.
horizontal) above 1 GeV in the \teq{\erg_p^2n_p(\erg_p)}
representation.  Deviations from a power-law behaviour can be up to
a factor of around 3 over four decades in particle energy in the 1
GeV -- 10 TeV range.  Such enhancements are extremely important for
discussion of the potential observability of EGRET sources by Whipple,
HEGRA and other ground-based TeV experiments.

\subsection{The maximum particle energy}

The maximum energy of Fermi-accelerated protons is determined by two
considerations, namely the acceleration time vs. the age of the
remnant, and the diffusive scalelength of the particles vs. the size of
the ionized region.  For test-particle shocks (i.e. linear ones that
produce power-law protons), the acceleration timescale (e.g. see Forman
and Morfill 1979, Drury 1983 for derivations) for the Fermi mechanism
at plane-parallel shocks up to energy \teq{E_{\rm TeV}} (in units of
TeV), far above the injection energy, can be written as
\begin{equation}
   \tau_{\rm a}\; =\; 31.7\,\dover{\eta}{QB_{-5}}\,
   \dover{r(r+1)}{r-1}\; \dover{E_{\rm TeV}}{u_{1000}^2}\;\;\;\hbox{yr},
\end{equation}
where \teq{r=u_{\rm ups}/u_{\rm down}} is the compression ratio,
\teq{B_{-5}} is the field in units of \teq{10^{-5}}Gauss, and
\teq{\eta} is the ratio of the particle mean free path \teq{\lambda} to
its gyroradius \teq{r_g}.  Here \teq{Q} is the charge number of the ion
(e.g. \teq{Q=1} for protons and \teq{Q=2} for alpha particles).  Eq.~(1)
is a specialization of Eq.~(4) of Ellison, Baring and Jones (1995),
expressed in units appropriate to the SNR problem at hand.  Quite
generally, \teq{\eta\geq 1} and one infers values of \teq{1 <\eta <10}
in interplanetary shocks in the inner heliosphere (Baring et al.
1997).  Shock speeds \teq{u_{1000}} (\teq{=u_{\rm ups}}) are in units
of 1000 km/s.  Implicit in such a formula is the assumption that the
mean free path scales as \teq{\lambda\propto r_g\propto p}, which seems
reasonable in the light of the aforementioned observations of plasmas
in the heliosphere.  Although Eq.~(1) is strictly confined to
test-particle regimes, it serves also as an order-of-magnitude estimate
for the full non-linear problem that is treated by the Monte Carlo
simulation.  Assume that the age of remnants is just their shell radii,
typically in the vicinity of a few pc, divided by the expansion speed
\teq{u_{1000}}.  This contention can be modified by the dynamical
interaction of the supernova ejecta with the ISM, leading to larger age
estimates (e.g.  Chevalier 1982).  For SN Ia, typically
\teq{u_{1000}\sim 10}, while for the more-massive ejecta associated
with SN II, the shock speed is slower:  \teq{u_{1000}\sim 4}.  Hence
(bright) remnant ages are typically \teq{1-10\times 10^{10}}sec,
implying age-limited acceleration terminating at around
\teq{10^4-10^5}GeV.

The diffusion scale of the SNR medium is \teq{\kappa /u} for a
diffusion coefficient of \teq{\kappa =\lambda v/3}.  Hence the
diffusion scale (in parsecs) is comparable to
\begin{equation}
   d_{\rm pc}\; =\; 3\times 10^{-2}\,\dover{\eta}{QB_{-5}}\,
   \dover{E_{\rm TeV}}{u_{1000}}\;\; .
\end{equation}
For acceleration regions of the order of a tenth of the shell size
(much smaller might be expected for remnants in the neighbourhood of
dense neutral regions, which is frequently the case for remnants
associated with unidentified EGRET sources), size-limited acceleration
terminates also at around \teq{10^4-10^5}GeV.  The factor of a tenth
represents an estimate of the effective escape lengthscale taking
into account shell geometry and the remnant history.  Note that if
the acceleration time in Eq.~(1) is multiplied by the shock speed, the
resulting length scale is of the order of Eq.~(2), differing only by
factors involving the compression ratio.  This is not surprising since
both estimates have their origin in particle diffusion.  Clearly higher
compression ratios will enhance the acceleration time, defining the
general property that stronger shocks usually produce size-limited
rather than time-limited  maximum energies of acceleration.  Another
important property of these estimates is that Eqs.~(1) and ~(2) depend
on the charge of the species but not explicitly on the mass.  Hence
electrons potentially have similar maximum energies to protons if
cooling is absent, while helium and heavier nuclei with higher charge
states can be accelerated to higher energies (as opposed to energy per
nucleon).

\section{GAMMA-RAY EMISSION SPECTRA}

The effects of using the full non-linear Monte Carlo simulation for
predicting gamma-ray emissivities from SNRs can be adequately
demonstrated using just the \teq{\pi^0} emission component.  Treatment
of other radiation mechanisms (discussed below) is deferred to future
work, in part because the hadronic contributions tend to provide the
major portion of the gamma-ray radiation.  The \teq{\pi^0} emissivity
and spectrum were calculated much along the lines of the work of Dermer
(1986): \teq{pp} collisions produce \teq{\pi^0}s, which subsequently
decay to produce two photons.  The proton component of the cosmic rays
collides with nuclei in the cold ambient ISM; note that cosmic rays of
higher mass number also contribute significantly to the \teq{\pi^0}
emissivity.  Low-energy protons, typically with kinetic energies below
a few GeV, create pions via the \teq{\Delta} resonance, following the
model of Stecker (1970).  As their energy increases, more resonances
are sampled.  High energy protons, typically above 10 GeV, interact
according to a radial scaling empirical formalism (e.g. Tan and Ng
1983).  Details of these formalisms, kinematics and numerical
procedure, can be found in Baring and Stecker (1997, in preparation).
Note that the gamma-ray spectral profiles for different cosmic ray
proton kinetic energies are symmetric about the energy
\teq{m_{\pi}c^2/2}, due to the pion decay kinematics (e.g. Stecker
1970).

Fig.~2 shows various representations of a \teq{\pi^0} gamma-ray
spectrum resulting from a proton distribution generated by the
non-linear Monte Carlo simulation.  In this case, protons are
accelerated out to a few tens of TeV, a limit obtainable from Eq.~(1)
for a very young remnant with \teq{\eta\sim 10} and \teq{r\sim 11} (it
must be remembered that non-linear shocks generate overall compression
ratios much larger than 4; see Jones and Ellison 1991; Ellison, Baring
and Jones 1996).  The differential photon spectrum is largely visually
uninteresting, so integral and double-integral (``\teq{\nu F_{\nu}}'')
spectra are plotted to highlight the spectral properties.  One
prominent feature is that spectra flatter than \teq{\erg_{\gamma}^{-2}}
result.  The spread induced by the kinematic phase space for pion
production and decay tends to smear out the underlying curvature of the
proton population.  Yet, for comparison, the integral spectrum
resulting from a power-law proton population with a high-energy cutoff
is depicted in Fig.~2, indicating that it overestimates the TeV/EGRET
flux ratio by a factor of a few.  This clearly yields an indication of
the improvement of the non-linear calculation over the standard
test-particle \teq{p^{-2}} infinite power-law case considered by Drury,
Aharonian and V\"olk (1994, and also approximately by Gaisser,
Protheroe and Stanev 1996).  Therefore it follows that these
test-particle implementations of shock acceleration theory can
significantly overpredict the TeV/EGRET flux ratio.  Consequently, the
incorporation of non-linear dynamical considerations can potentially
relax observational constraints imposed by the Whipple upper limits
to IC443, W28 and $\gamma$-Cygni as presented by Lessard et al.
(1995).  Note that the spectral structure below 1 GeV in the proton
distribution in Fig.~1 is immaterial since such kinetic energies are
below the \teq{\pi^0} production threshold.

%
\vskip+0.5truecm
\centerline{\hskip -0.0truecm\psfig{figure=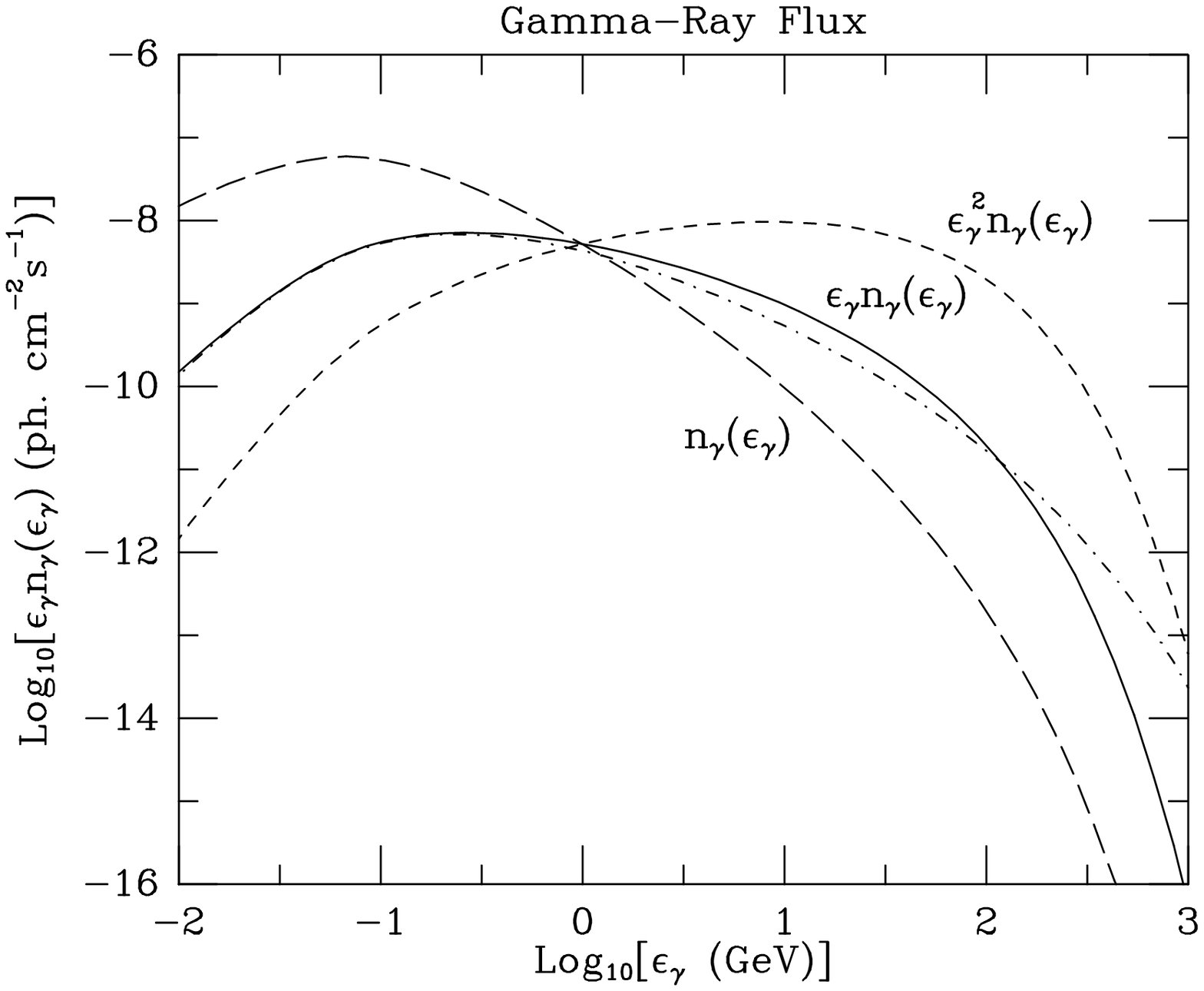,width=7.9truecm}}
\vskip-0.3truecm
{\smallskip \it Figure~2:  The $\gamma$-ray flux at earth, for a young
SNR about a kpc distant, integrated over a shock acceleration produced
proton distribution (long dashed lines).  The solid curve represents
the integral spectrum for this distribution, and the dash-dot curve the
integral spectrum from a pure power-law (\teq{p^{-2}}) proton
population (exponentially cutoff at around 1 TeV) with index given by
the low energy portion of the non-linear model proton distribution.
The \teq{\erg_{\gamma}^2n_{\gamma}} representation clearly illustrates
the effects of spectral curvature and the cutoff.   \smallskip}

DeJager and Mastichiadis (1996) suggested that constraints to the
TeV/EGRET flux ratio can be relieved by having the photon spectrum turn
over below the Whipple sensitivity threshold.  The maximum energies of
the protons at \teq{\sim 10-30} TeV propagates down into the few TeV
range in the gamma-ray spectrum due to the inelasticity of the
\teq{pp\to pp\pi^0\to pp\gamma\gamma} chain.  Hence for the illustrated
case, HEGRA would be unlikely to see an EGRET source, but Whipple
potentially could.  Turnovers at such low energies are predicted
neither in the work of Gaisser, Protheroe and Stanev (1996), nor by
Drury, Aharonian and V\"olk (1994), mainly because they focus on
cooling-limited rather than age/size-limited acceleration.  Their
inverse Compton cooling rates off the microwave and local IR
backgrounds turn out to be low enough that the freely expanding shell
purely defines the maximum proton energy.  The non-linear shock
acceleration solution treated here provides natural cutoffs in the TeV
range that can also yield compatibility with the Whipple upper limits
for the appropriate choice of remnant shock parameters.  By inspection
of Eq.~(2), slightly lower shock speeds can suppress emission in the
TeV range, which may well be necessary to explain the four orders of
magnitude difference in integral flux inferred for $\gamma$-Cygni (see
Lessard et al. 1995) between 100 MeV and 500 GeV from the EGRET source
flux and the Whipple upper limits.  The implication of such low maximum
energies for cosmic ray protons is naturally that older SNRs are
required to produce the bulk of cosmic rays out to the \teq{10^{14}eV}
``knee,'' so that gamma-ray SNRs are a gamma-ray bright minority of the
cosmic ray-producing remnant population.

\section{DISCUSSION}

Generalizing beyond \teq{\pi^0} emission generated via \teq{pp}
collisions is clearly important.  These interactions also spawn charged
pions at comparable rates, and these yield secondary electrons and
positrons and consequently inverse Compton emission as discussed in
Gaisser, Protheroe and Stanev (1996).  While the \teq{\pi^0} decay
spectrum traces the proton distribution, the inverse Compton spectrum
is potentially much flatter due to the kinematics.  Hence, in
principal, the inverse Compton mechanism can be much more efficient at
higher energies, and perhaps dominate the \teq{\pi^0} decay products;
such scenarios are discussed in the context of the diffuse galactic
gamma-ray background by Hunter et al. (1996).  Primary
shock-accelerated electrons are in principal possible, though the
theoretical efficiency of their generation is very poorly understood.
Gaisser, Protheroe and Stanev (1996) argue that the EGRET/Whipple
spectral constraints for IC 443 imply that inverse Compton emission is
relatively unimportant and also that the primary electron population is
of low abundance.  Within the confines of non-linear shock acceleration
theory, unlike the power-law choices made (i.e. linear shock models) by
Gaisser, Protheroe and Stanev (1996) and De Jager and Mastichiadis
(1996), the accelerated electron spectrum does not trace the proton
distribution at all energies.  Below a few GeV, the electron diffusion
lengths are always less than those of protons of comparable energy, so
that electrons effectively experience a weaker portion of the shock
structure than do the protons (e.g. see Ellison and Reynolds 1991 for a
discussion).  The net effect is that the electron distributions are
steeper below a few GeV and the population level above a few GeV much
lower than is predicted in the linear shock models.  Therein lies
another crucial aspect of the inclusion of the non-linear effects of
shock acceleration theory: the linear approaches cited above seriously
overestimate the efficiency of electron acceleration, and hence the
contribution of the inverse Compton and bremsstrahlung processes to the
SNR luminosity.  Such issues will be dealt with in a future paper
(Baring et al. 1997, in preparation).

The issue of the prominence or otherwise of an inverse Compton
component may well be elucidated by coordinated observations at the TeV
range and by INTEGRAL at MeV energies: the \teq{\pi^0} decay channel
produces much lower emissivity below \teq{m_{\pi}c^2/2\sim 67}MeV due
to decay kinematics, so that an inverse Compton spectrum can be much
more visible in the soft gamma-ray regime.  An asset of INTEGRAL will
be its angular-resolution capability (e.g. see Winkler
1996).  EGRET has difficulty resolving on scales of the order of the
angular diameter of nearby remnants; hence it is not clear whether any
detected gamma-ray emission is associated with whole remnants, portions
or ``hot spots'' (or neither).  Improved angular resolution at MeV and
TeV energies will clearly pin down or disprove these associations.
Note also that, apart from protons, heavier nuclei make up the cosmic
ray population and the ambient ISM, and these also contribute
significantly to the gamma-ray and electronic products; their inclusion
in our gamma-ray emissivity calculations is deferred to future work.
In conclusion, coupled TeV/sub-GeV observations of supernova remnants,
in which the INTEGRAL can play a big role, will discriminate between
various gamma-ray emission models of these sources, and enhance our
understanding of the cosmic ray production.  The non-linear effects
described here are an essential ingredient for any model that invokes
Fermi acceleration at SNRs, particularly since their prediction of
enhanced TeV/GeV flux ratios can tighten model constraints imposed by
current TeV upper limits and possible future detections.

\section*{ACKNOWLEDGMENTS}

We thank S. Reynolds and P. Goret for many helpful and informative 
discussions.



\clearpage



\end{document}